\documentclass[aps,prl,superscriptaddress,showpacs,floatfix,twocolumn]{revtex4}

\usepackage{graphicx}

\def\v2{\mbox{$v_2$}}

\def\sqrtsNN{\mbox{$\sqrt{s_{NN}}$}}

\begin{document}


\title{Evidence for a long-range component in the pion emission source in
Au+Au collisions at \sqrtsNN=200~GeV }

\newcommand{\abilene}{Abilene Christian University, Abilene, TX 79699, U.S.}
\newcommand{\acadsin}{Institute of Physics, Academia Sinica, Taipei 11529, Taiwan}
\newcommand{\banaras}{Department of Physics, Banaras Hindu University, Varanasi 221005, India}
\newcommand{\barc}{Bhabha Atomic Research Centre, Bombay 400 085, India}
\newcommand{\bnl}{Brookhaven National Laboratory, Upton, NY 11973-5000, U.S.}
\newcommand{\caucr}{University of California - Riverside, Riverside, CA 92521, U.S.}
\newcommand{\ciae}{China Institute of Atomic Energy (CIAE), Beijing, People's Republic of China}
\newcommand{\cns}{Center for Nuclear Study, Graduate School of Science, University of Tokyo, 7-3-1 Hongo, Bunkyo, Tokyo 113-0033, Japan}
\newcommand{\columbia}{Columbia University, New York, NY 10027 and Nevis Laboratories, Irvington, NY 10533, U.S.}
\newcommand{\dapnia}{Dapnia, CEA Saclay, F-91191, Gif-sur-Yvette, France}
\newcommand{\debrecen}{Debrecen University, H-4010 Debrecen, Egyetem t{\'e}r 1, Hungary}
\newcommand{\elte}{ELTE, E{\"o}tv{\"o}s Lor{\'a}nd University, H - 1117 Budapest, P{\'a}zm{\'a}ny P. s. 1/A, Hungary}
\newcommand{\fsu}{Florida State University, Tallahassee, FL 32306, U.S.}
\newcommand{\gsu}{Georgia State University, Atlanta, GA 30303, U.S.}
\newcommand{\hiroshima}{Hiroshima University, Kagamiyama, Higashi-Hiroshima 739-8526, Japan}
\newcommand{\ihepprot}{IHEP Protvino, State Research Center of Russian Federation, Institute for High Energy Physics, Protvino, 142281, Russia}
\newcommand{\isu}{Iowa State University, Ames, IA 50011, U.S.}
\newcommand{\jinrdubna}{Joint Institute for Nuclear Research, 141980 Dubna, Moscow Region, Russia}
\newcommand{\kaeri}{KAERI, Cyclotron Application Laboratory, Seoul, South Korea}
\newcommand{\kangnung}{Kangnung National University, Kangnung 210-702, South Korea}
\newcommand{\kek}{KEK, High Energy Accelerator Research Organization, Tsukuba, Ibaraki 305-0801, Japan}
\newcommand{\kfki}{KFKI Research Institute for Particle and Nuclear Physics of the Hungarian Academy of Sciences (MTA KFKI RMKI), H-1525 Budapest 114, POBox 49, Budapest, Hungary}
\newcommand{\korea}{Korea University, Seoul, 136-701, Korea}
\newcommand{\kurchatov}{Russian Research Center ``Kurchatov Institute", Moscow, Russia}
\newcommand{\kyoto}{Kyoto University, Kyoto 606-8502, Japan}
\newcommand{\labllr}{Laboratoire Leprince-Ringuet, Ecole Polytechnique, CNRS-IN2P3, Route de Saclay, F-91128, Palaiseau, France}
\newcommand{\lawllnl}{Lawrence Livermore National Laboratory, Livermore, CA 94550, U.S.}
\newcommand{\losalamos}{Los Alamos National Laboratory, Los Alamos, NM 87545, U.S.}
\newcommand{\lpc}{LPC, Universit{\'e} Blaise Pascal, CNRS-IN2P3, Clermont-Fd, 63177 Aubiere Cedex, France}
\newcommand{\lund}{Department of Physics, Lund University, Box 118, SE-221 00 Lund, Sweden}
\newcommand{\muenster}{Institut f\"ur Kernphysik, University of Muenster, D-48149 Muenster, Germany}
\newcommand{\myongji}{Myongji University, Yongin, Kyonggido 449-728, Korea}
\newcommand{\nagasaki}{Nagasaki Institute of Applied Science, Nagasaki-shi, Nagasaki 851-0193, Japan}
\newcommand{\newmex}{University of New Mexico, Albuquerque, NM 87131, U.S.}
\newcommand{\nmsu}{New Mexico State University, Las Cruces, NM 88003, U.S.}
\newcommand{\ornl}{Oak Ridge National Laboratory, Oak Ridge, TN 37831, U.S.}
\newcommand{\orsay}{IPN-Orsay, Universite Paris Sud, CNRS-IN2P3, BP1, F-91406, Orsay, France}
\newcommand{\pnpi}{PNPI, Petersburg Nuclear Physics Institute, Gatchina,  Leningrad region, 188300, Russia}
\newcommand{\riken}{RIKEN, The Institute of Physical and Chemical Research, Wako, Saitama 351-0198, Japan}
\newcommand{\rikjrbrc}{RIKEN BNL Research Center, Brookhaven National Laboratory, Upton, NY 11973-5000, U.S.}
\newcommand{\saispbstu}{Saint Petersburg State Polytechnic University, St. Petersburg, Russia}
\newcommand{\saopaulo}{Universidade de S{\~a}o Paulo, Instituto de F\'{\i}sica, Caixa Postal 66318, S{\~a}o Paulo CEP05315-970, Brazil}
\newcommand{\seoulnat}{System Electronics Laboratory, Seoul National University, Seoul, South Korea}
\newcommand{\stonybrkc}{Chemistry Department, Stony Brook University, SUNY, Stony Brook, NY 11794-3400, U.S.}
\newcommand{\stonycrkp}{Department of Physics and Astronomy, Stony Brook University, SUNY, Stony Brook, NY 11794, U.S.}
\newcommand{\subatech}{SUBATECH (Ecole des Mines de Nantes, CNRS-IN2P3, Universit{\'e} de Nantes) BP 20722 - 44307, Nantes, France}
\newcommand{\tenn}{University of Tennessee, Knoxville, TN 37996, U.S.}
\newcommand{\titech}{Department of Physics, Tokyo Institute of Technology, Oh-okayama, Meguro, Tokyo, 152-8551, Japan}
\newcommand{\tsukuba}{Institute of Physics, University of Tsukuba, Tsukuba, Ibaraki 305, Japan}
\newcommand{\vandy}{Vanderbilt University, Nashville, TN 37235, U.S.}
\newcommand{\waseda}{Waseda University, Advanced Research Institute for Science and Engineering, 17 Kikui-cho, Shinjuku-ku, Tokyo 162-0044, Japan}
\newcommand{\weizmann}{Weizmann Institute, Rehovot 76100, Israel}
\newcommand{\yonsei}{Yonsei University, IPAP, Seoul 120-749, Korea}
\affiliation{\abilene}
\affiliation{\acadsin}
\affiliation{\banaras}
\affiliation{\barc}
\affiliation{\bnl}
\affiliation{\caucr}
\affiliation{\ciae}
\affiliation{\cns}
\affiliation{\columbia}
\affiliation{\dapnia}
\affiliation{\debrecen}
\affiliation{\elte}
\affiliation{\fsu}
\affiliation{\gsu}
\affiliation{\hiroshima}
\affiliation{\ihepprot}
\affiliation{\isu}
\affiliation{\jinrdubna}
\affiliation{\kaeri}
\affiliation{\kangnung}
\affiliation{\kek}
\affiliation{\kfki}
\affiliation{\korea}
\affiliation{\kurchatov}
\affiliation{\kyoto}
\affiliation{\labllr}
\affiliation{\lawllnl}
\affiliation{\losalamos}
\affiliation{\lpc}
\affiliation{\lund}
\affiliation{\muenster}
\affiliation{\myongji}
\affiliation{\nagasaki}
\affiliation{\newmex}
\affiliation{\nmsu}
\affiliation{\ornl}
\affiliation{\orsay}
\affiliation{\pnpi}
\affiliation{\riken}
\affiliation{\rikjrbrc}
\affiliation{\saispbstu}
\affiliation{\saopaulo}
\affiliation{\seoulnat}
\affiliation{\stonybrkc}
\affiliation{\stonycrkp}
\affiliation{\subatech}
\affiliation{\tenn}
\affiliation{\titech}
\affiliation{\tsukuba}
\affiliation{\vandy}
\affiliation{\waseda}
\affiliation{\weizmann}
\affiliation{\yonsei}
\author{S.S.~Adler}	\affiliation{\bnl}
\author{S.~Afanasiev}	\affiliation{\jinrdubna}
\author{C.~Aidala}	\affiliation{\bnl}
\author{N.N.~Ajitanand}	\affiliation{\stonybrkc}
\author{Y.~Akiba}	\affiliation{\kek} \affiliation{\riken}
\author{J.~Alexander}	\affiliation{\stonybrkc}
\author{R.~Amirikas}	\affiliation{\fsu}
\author{L.~Aphecetche}	\affiliation{\subatech}
\author{S.H.~Aronson}	\affiliation{\bnl}
\author{R.~Averbeck}	\affiliation{\stonycrkp}
% ...
\collaboration{PHENIX Collaboration} \noaffiliation

\date{\today}

\begin{abstract}
Emission source functions are extracted from correlation functions
constructed from charged pions produced at mid-rapidity in Au+Au collisions
at \sqrtsNN=200~GeV. The source parameters extracted from these functions at
low $k_T$, give first indications of a long tail for the pion emission
source. The source extension cannot be explained solely by simple kinematic
considerations. The possible role of a halo of secondary pions from
resonance emissions is explored.
\end{abstract}


\maketitle

Collisions between ultra-relativistic heavy ions can lead to extremely high 
energy-density nuclear matter~\cite{Adcox:2004mh}. 
The decay dynamics of this matter are strongly influenced by 
the nuclear Equation of State (EOS) and possibly by a de-confined 
phase~\cite{Shuryak:2004cy}. 
An emitting system which undergoes a strong first order phase transition is 
expected to show a much larger space-time extent than would be expected if the 
system remained in the hadronic phase 
throughout the collision process \cite{Pratt:1984su}.
Indeed, several hydrodynamical calculations show  
such an increase 
for particle emission sources \cite{Pratt:1984su,Teaney:2000cw}, 
providing hadronization does not occur via a supercooled state \cite{Csorgo:1994dd}.
It has also been suggested that the shape of the emission source function can 
provide signals for a second order phase transition and whether or not particle 
emission occurs near to the critical end point in the QCD phase diagram \cite{Csorgo:2005it}.

Interferometry studies provide important information on the 
emission source function for particles produced in nuclear 
reactions ranging from elementary collisions ($e^+ e^-$ 
and $(\overline{p})pp$) to those involving 
very 
heavy ions~\cite{Shuryak:1972kq,Heinz:1999rw,Csorgo:1999sj}. 
Recent studies include the use of two- and three-pion interferometry correlations 
spanning the beam energies of the Alternating Gradient Synchrotron (AGS), 
the Super Proton Synchrotron (SPS) and the Relativistic Heavy Ion Collider 
(RHIC) (\sqrtsNN $\sim 2-200$~GeV) \cite{Lisa:2005dd,
Adler:2001zd,Adcox:2002uc,Adams:2003vd}. 

	A common theme for these papers is the extraction of the widths of
emission source functions which are assumed to be Gaussian. For such
extractions, the Coulomb effects on the correlation function are usually
assumed to be separable \cite{Sinyukov:1998fc} as well.
Such an approach was followed in an earlier analysis in which we used the    
Bowler-Sinyukov 3D HBT analysis method [in Bertsch-Pratt 
coordinates] to probe for long-range emissions from a possible long-lived 
source \cite{Adler:2004rq}. The RMS-widths so obtained for each dimension of 
the source $R_{\text{long}}, R_{\text{side}}$ and $R_{\text{out}}$, gave no 
evidence for such emissions, suggesting that the sound speed is not 
zero during an extended hadronization period. 

In this letter we exploit the model-independent imaging technique of Brown and 
Danielewicz~\cite{Brown:1997ku,Brown:2000aj} to make a more detailed study of  
both the shape and the space-time characteristics of the pion emission source 
function. The method uses a numerical calculation of the two particle wave function 
(including final state interactions (FSI)) to produce an inversion matrix 
that operates on the correlation function to produce the corresponding 
source function. 
The technique has been used to address only a few data 
sets~\cite{Panitkin:2001qb,Chung:2002vk} at relativistic beam energies.

The measurements were made with the PHENIX detector \cite{Adcox:2003zm} 
at RHIC. 
The charged pions relevant to this analysis were detected in the two central 
arms ($|\eta|\leq0.35$). 
Track reconstruction was accomplished via pattern 
recognition using the drift chamber (DC) followed by two layers of multi-wire 
proportional chambers with pad readout 
located at radii of 2, 2.5, and 5~m~\cite{Adcox:2003zm}.
Particle momenta were measured with the resolution $\delta$p/p~=~0.7\%~$\oplus 1.0$\%p~(GeV/$c$). 
Very good pion identification (PID) was achieved with a $2 \sigma$ cut about the 
pion peak in the squared-mass distribution for $p_T \lesssim 2.0$~GeV/$c$  
and $p_T \lesssim 1$~GeV/$c$ in the Time of Flight (TOF) and the 
Electromagnetic Calorimeter (EMC) respectively.
The event centrality was determined using the PHENIX beam-beam 
counter and the zero degree calorimeter \cite{Adcox:2003nr}.
%
%
Approximately 22 million Au+Au events were analyzed to generate and study correlation 
functions for several centrality and $p_T$ selections.

	Two-pion interferometry correlations were obtained via the correlation function
$C(q)= {N_{cor}(q)}/{N_{mix}(q)}$, where the numerator is the relative momentum 
distribution of particle pairs from 
the same event (foreground pairs) and the denominator is the relative momentum distribution 
of particle pairs obtained from mixed events (background pairs). 
Here, $q=\frac{1}{2}\sqrt{-(p_1-p_2)^2}$ is half of the
relative momentum between the two particles in the pair c.m. frame (PCMS).  $p_1$ and 
$p_2$ are the momentum 4-vectors of each particle in the pair and $C(q) \equiv 1$  
for large $q$-values where final-state interactions are negligible. 
Track-pair cuts similar to those of Ref.~\cite{Adcox:2002uc} were applied to 
foreground and background pairs respectively. That is,  
pairs within 0-5~cm in the beam direction ($\Delta Z_{DC}$) and
0.02 radians in azimuthal angle ($\Delta\phi_{DC}$) in the DC were
eliminated from the pair sample to remove ghost tracks, and pairs 
within $0.0 < \Delta\phi_{DC} <0.1$ radians for $\Delta Z_{DC} >5$~cm
were removed to avoid an inefficient region. The latter 
set of cuts were supplemented with the removal of pairs having a 
separation $\Delta R \le$ 14 cm and $\le 16$ cm in the TOF and EMC 
respectively. Systematic variations of all of these cuts were explored 
to determine systematic error estimates; very little influence on the 
extracted correlation functions was found. 
Careful studies of the influence of momentum resolution on the correlation 
function were also made following the simulation technique  
outlined in Ref.~\cite{Adcox:2002uc}. The maximum effect was found to 
be $\sim 0.4$\% change in the correlation function at low $q$ values, 
so it was neglected.

\begin{figure}[tbh]
\includegraphics[width=1.0\linewidth]{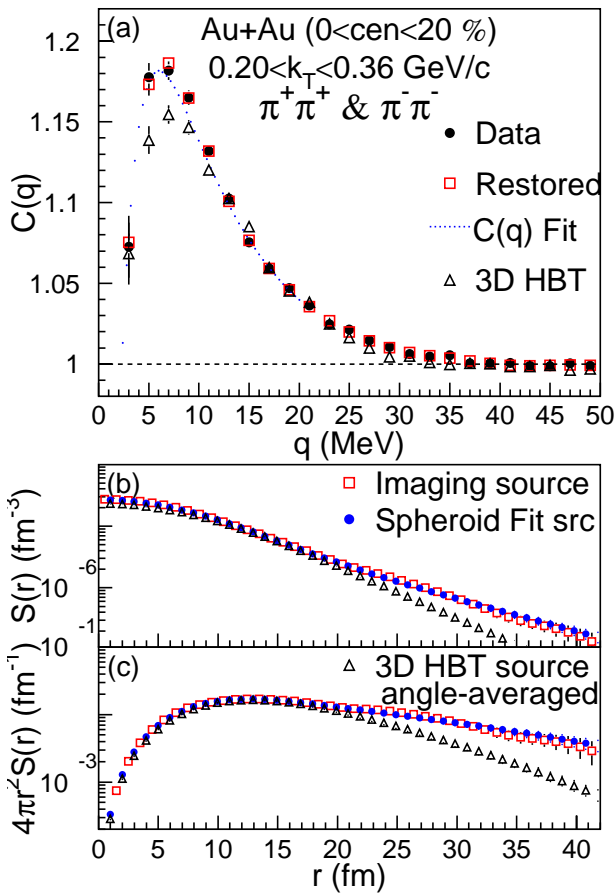}
\caption{\label{fig1} (color online)
(upper, a) (filled circles) Correlation function, C(q) 
for $\pi^+\pi^+$ and $\pi^-\pi^-$ pairs;  
(open squares) restored correlation from imaging technique; 
(dotted line) direct correlation fitting; 
(open triangles) 1D angle-averaged correlation of 3D correlation function.  
(lower) 1D source function (b) S(r) and (c) 4$\pi r^2$S(r): 
(open squares) imaging; 
(filled circles) spheroid fit to correlation function; 
(open triangles) angle-averaging of 3D-Gaussian source function. 
Systematic errors are less than size of data points.
}
\end{figure}

	The filled circles in Fig.~\ref{fig1}(a) show the  one-dimensional 
(1D) correlation function $C(q)$, for a centrality of 
0 - 20\% and for $0.20 < k_T = \frac{1}{2}(p_{T,1} + p_{T,2}) <0.36$~GeV/$c$. 
The characteristic enhancement for $q \alt 25$~MeV/c reflects a combination of 
Bose-Einstein statistics and the FSI between pion pairs. 
The correlation function is not Coulomb corrected because the FSI 
(including Coulomb but no strong interaction) 
is included in the imaging and fitting procedure as described below. 

%
	The 1D correlation function and source function $S(r)$ 
(the probability of emitting a pair of particles at a separation $r$ in 
the PCMS frame), are related  via the 1D~Koonin-Pratt 
equation \cite{Koonin:1977fh}:
%
%
\begin{equation}
  C(q)-1 = 4\pi\int dr r^2 K_0(q,r) S(r).
%
%
\label{kpeqn}
\end{equation}
The angle-averaged kernel $K_0(q,r)$ encodes the FSI and is given in terms of the 
final state wave function $\Phi_{\bf q}({\bf r})$, as $K_0(q,r)=\frac{1}{2}\int
d(\cos(\theta_{\bf q, r})) (|\Phi_{\bf q}({\bf r})|^2-1)$, where $\theta_{\bf q, r}$ 
is the angle between {\bf q} and {\bf r}~\cite{Brown:2000aj}. The procedure 
for the inversion of Eq.~\ref{kpeqn} to obtain $S(r)$ is also given in 
Ref.~\cite{Brown:2000aj}. 
%
%
%
%

	The open squares in Fig.~\ref{fig1}(b) show the source function
obtained from the correlation function presented in Fig.~\ref{fig1}(a). As
a cross check of the imaging procedure, a restored correlation function
was generated via Eq.~\ref{kpeqn} with the extracted source function as
input. The open squares and filled circles in Fig.~\ref{fig1}(a)  
indicate excellent consistency between the measured and restored
correlation functions.  The 1D-source function (cf. Fig.~\ref{fig1}(b))
points to a Gaussian-like pattern at small $r$ and a previously unresolved
``tail" at large $r$.  The robustness of this tail was established via an
extensive array of tests including its dependence on pair and PID cuts,
and on momentum resolution; no variation outside of the stated error bars
was found.

	This new observation of a tail is made more transparent via a 
comparison with the source function constructed from 
the parameters ($R_{\text{long}}, R_{\text{side}}$, $R_{\text{out}}$ 
and $\lambda$), obtained in an earlier 3D HBT analysis \cite{Adler:2004rq}.
The procedures outlined in Ref.\cite{Brown:2000aj} were employed to construct 
this source function (see Fig.~\ref{fig1}). 
The measured and 3D angle-averaged correlation 
functions differ for $q \alt 15$~MeV/c
as do the respective source functions for $r \agt 17$~fm. 
The imaged source function exhibits a more prominent 
tail than the angle averaged 3D HBT source function. 
This difference could stem from the Gaussian shape assumption 
employed in the 3D HBT analysis. 
The 3D Gaussian fitting procedure by construction is sensitive 
only to the main component of S(r), and thus would not be capable 
of resolving fine structure at small-q/large-r.
Given the fact that the volume element increases 
quadratically with pair separation, this difference is considerable as 
shown in Fig.\ref{fig1}(c), where the radial probabilities (4$\pi r^2$S(r)) 
are compared. The open triangles in Fig.~\ref{fig1}(a) clearly indicate that 
the differences in the source functions reflect an important disparity 
in the corresponding correlation functions for $q \alt 10~-~15$~MeV.
%

	Parameters of the source function for different assumed shapes 
were extracted via direct fits to the correlation function. 
Filled circles in Figs.~\ref{fig1}(b,c) show the source 
function obtained from such a fit for a Spheroidal shape~\cite{Brown:2000aj}, 
%
%
\begin{equation}
S(r) = \frac{\lambda \: R_{\text{eff}} \times e^{ - \frac {r^2}{4R_T^2} } 
\text{erfi}(\frac{r}{2 R_{\text{eff}} }) }{\left( \text{8} \pi R_T^2 R_0 r \right)}, \; \text{for} \; R_0 > R_T, 
%
\label{blimp}
\end{equation}
%
%
where $R_{\text{eff}} = 1/ \sqrt{(1/{R_T^2} - 1/{R_0^2}) }$, $R_{T}$ is the 
radius of the Spheroid in two perpendicular spatial dimensions and 
$R_0 = a \times R_T$ is the radius in the third spatial dimension; 
$a$ is a scale factor.  
The long axis of the Spheroid is assumed to be oriented in the out 
direction of the Bertsch-Pratt coordinate system.
The fraction of pion pairs which contribute to the source $\lambda$,
is given by the integral of the normalized source function over the 
full range of $r$. 

	The procedure for making a direct fit to the correlation function involves the 
determination of a set of values for the Spheroid parameters of Eq.~\ref{blimp}, which 
reproduce the observed correlation function when the resulting source function is 
inserted into Eq.~\ref{kpeqn}. 
The minimization package MINUIT was used to minimize the $\chi^2$ between the 
observed and calculated correlation function. The $\chi^2$/ndf value so obtained 
was $\sim 1$. The dotted curve in Fig.~\ref{fig1}(a) shows the fit to the data. 
The resulting source function shown in Figs.~\ref{fig1}(b) 
and \ref{fig1}(c), indicates very good agreement with that obtained via the 
imaging technique. 
This shape parametrization is not unique; an essentially indistinguishable 
source was also obtained 
for a fit performed with a Gaussian plus modified 
exponential \cite{Chung:2005ra} shape. The simpler 
Spheroid parametrization was chosen for the discussion below.

\begin{figure}[tb]
\includegraphics[width=1.0\linewidth]{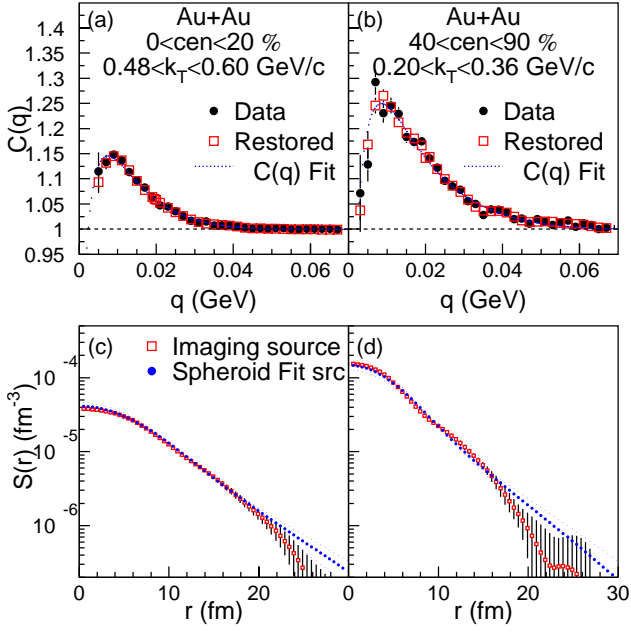}

\caption{\label{fig2}  (color online)
(upper) Correlation functions, C(q), for $\pi^+\pi^+$ and $\pi^-\pi^-$ pairs 
and (lower) corresponding source functions, S(r) for
(a, c) high $k_T$ most central collisions;
(b, d) low $k_T$ peripheral collisions.
Error bars indicate statistical errors with symbols as 
in Fig.~\protect\ref{fig1}.
}
\end{figure}

	The Spheroidal source function (Eq.~\ref{blimp}) can be
approximated by a short-range Gaussian source $S_{sr}(r)$;
%
%
\begin{equation}
S_{sr} (r) \sim \lambda\: e^{ - r^2\left[{\frac{1}{6R_T^2}} + {\frac{1}{12R_0^2}} \right] } /\left( {8\pi \sqrt \pi  R_T^2 R_0 } \right),
%
\label{Sr_s}
\end{equation}
for small $r$, and a long-range source $S_{lr}(r)$ for 
$r >> 2R_TR_0/\sqrt{R_0^2 - R_T^2}$ given by 
%
%
\begin{equation}
S_{lr} (r) \sim \lambda\: R_0 e^{-{r^2 / 4R_0^2 } } 
/\left( {4\pi \sqrt \pi  \left( {R_0^2  - R_T^2 } \right)r^2 } \right).
%
%
\label{Sr_l}
\end{equation}
Thus, the emission source shown in Figs.~\ref{fig1}(b)
and \ref{fig1}(c) can be interpreted to reflect a short range Gaussian 
source of radius $R_{sr} = \sqrt{3R_T^2R_0^2/(2R_0^2+R_T^2)}$ and a long-range tail of extended 
space-time extent $R_{lr} = R_0$.
The fraction of pairs associated with these sources 
$\lambda_{sr} = \lambda\: a^2(3/(2a^2 +1))^{3/2}$ and  
$\lambda_{lr} = (\lambda - \lambda_{sr})$, are obtained from 
Eq.~\ref{Sr_s} and the conservation of pairs respectively.

	Source functions were extracted via imaging and spheroid fits for 
several $k_T$ and centrality selections, in order to map 
the regions in $k_T$ and centrality where the long-range tail
is prominent. Representative results are shown in Fig.\ref{fig2} for 
the indicated cuts. The experimental and 
restored correlation functions, compared in Figs.\ref{fig2}(a,b), 
indicate good agreement consistent within the statistics, as 
do the corresponding source functions shown in Figs.\ref{fig2}(c,d).
%

\begin{figure}[tb]
%
%
\includegraphics[width=1.0\linewidth]{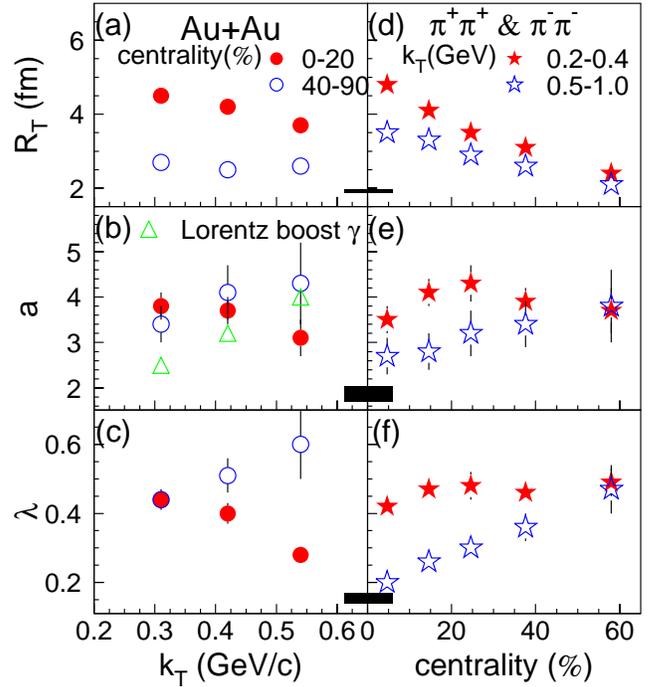} 
\caption{\label{fig3}  (color online)
(left) $k_T$ dependence of extracted spheroidal source parameters for 
pion source functions
(a) $R_T$, (b) $a$, and (c) $\lambda$ for
(filled circles) most central collisions;
(open circles) perhipheral collisions. 
(right) centrality dependence of same parameters for 
(filled stars) low $k_T$ ($0.2-0.4$~GeV/$c$);
(open stars) high $k_T$ ($0.5-1.0$ GeV/$c$).
Filled bands indicate systematic errors.
}
\end{figure}

Figure~\ref{fig3} gives a more complete summary of the extracted 
source parameters. 
The indicated systematic
errors were obtained by varying the pair cuts in the analysis.
The centrality and $k_T$ dependence of the RMS radius of the short-range 
source $R_{sr}$, is similar (within 10\%) to that obtained 
for $R_{\text{long}}$ and $R_{\text{side}}$ in an earlier 
analysis \cite{Adler:2004rq} (cf. Figs.~\ref{fig3} (a) and (d)). 
The long-range source shows an effective radius $R_{lr} = a\times R_T$, which 
is 3-4 times $R_T$ (cf. Figs.~\ref{fig3} (b) and (e)) with the largest 
values for low $k_T$ and the most central collisions. 
The fraction of pairs exhibiting these characteristics is given by the
$\lambda$ values shown in Figs.~\ref{fig3} (c, f); maximum prominence
is shown for low $k_T$ pairs from central collisions.

	A central question is the origin of the long-range contribution 
to the emitting source. Instantaneous freeze-out of an isotropic source in the 
longitudinal co-moving system (LCMS) would give $R_{lr} = R_{out} = \gamma\times R_T$ 
in the PCMS. Thus, the values for $a$ and $\gamma$, shown in Fig.~\ref{fig3}(b), can 
be directly compared. At low $k_T$, $\gamma$ is seen to be significantly 
less than $a$. Thus, a simple kinematic transformation from the LCMS to PCMS 
can not account for the observed source parameters at this $k_T$.  

Could a composite particle emission source comprised of a central core and 
a halo of long-lived resonances account for the observations?
For such emissions, the pairing 
between pions from the core and secondary pions from the halo is 
expected to dominate the long-range emissions \cite{Nickerson:1997js}. 
If it is assumed that this halo includes only $\omega$ decay ($c\tau \sim 24$~fm), 
one may compare the $\omega$ yield with a simple estimate of the fraction of 
pion pairs associated with the short- and long-range sources.
Using the $a$ and $\gamma$ values in Fig.~\ref{fig3}(b), 
$\lambda_{sr} \sim 0.22~-~0.32$, $\lambda_{lr} \sim 0.23~-~0.13$ and 
$\lambda_{lr}/\lambda_{sr} \sim 0.4~-~1.0$ for the lowest $k_T$.
The preliminary data for $\omega/\pi^- =0.1$ from 
Ref.~\cite{Broniowski:2004kg} gives an estimate for
$\lambda_{lr}/\lambda_{sr} \sim 2\times 0.1/\sqrt{\lambda_{sr}} \sim 0.35~-.43$ 
which lies at the lower end of the estimates obtained from the source parameters. 
Therefore, it is plausible
that a maximal kinematic boost 
combined with a halo of $\omega$s could account for $\lambda_{lr}/\lambda_{sr}$.
However, the steep centrality dependence of the radius of the long range 
source inferred from Figs.~\ref{fig3}(d,e) is not compatible 
with what would be expected for significant $\omega$ resonance contribution.
Also, initial comparisons to the results from a recent dynamical model 
calculation \cite{Kisiel:2006is} indicate that resonance emissions 
are not sufficient to fully account for the observed tails in the source 
functions.

In summary, we have made the first extraction of the full 
1D emission source function for pions in Au+Au collisions at 
RHIC (\sqrtsNN~=~200 GeV).
This source function points to separate but prominent contributions 
from short-range emissions and a long-range tail 
of larger space-time extent than has been previously observed. 
This tail cannot be explained solely by simple kinematic considerations 
associated with a frame transformation from the LCMS to the PCMS.
Further detailed 3D imaging measurements in conjunction 
with in depth model studies, are required to quantitatively 
pin down the origin of this tail and to determine whether or not 
it is related to a phase transition.

\ 
\ 
\ 
\ 


\end{document}